# Nearly massless Dirac fermions and strong Zeeman splitting in nodal-line semimetal ZrSiS probed by dHvA quantum oscillations

Jin Hu, Zhijie Tang, Jinyu Liu, Yanglin Zhu, Jiang Wei and Zhiqiang Mao[*]

Department of Physics and Engineering Physics, Tulane University, New Orleans, LA 70118, USA

Abstract

Topological semimetals represent a new class of quantum materials hosting Dirac/Weyl fermions. The essential properties of topological fermions can be revealed by quantum oscillations. Here we present the first systematic de Haas–van Alphen (dHvA) oscillation studies on the recently discovered topological Dirac nodal-line semimetal ZrSiS. From the angular dependence of dHvA oscillations, we have revealed the anisotropic Dirac bands in ZrSiS and found surprisingly strong Zeeman splitting at low magnetic fields. The Landé $g$-factor estimated from the separation of Zeeman splitting peaks is as large as 38. From the analyses of dHvA oscillations, we also revealed nearly zero effective mass and exceptionally high quantum mobility for Dirac fermions in ZrSiS. These results shed light on the nature of novel Dirac fermion physics of ZrSiS.

[*]email: zmao@tulane.edu

## I. Introduction

The recent discoveries of three-dimensional (3D) topological Dirac and Weyl semimetals [1-14] have stimulated enormous interest. These materials are characterized by symmetry-protected discrete band touching points in the momentum space, at which the electron energy bands display linear crossing near the Fermi level [1, 3, 7-8]. The exotic properties resulting from Dirac and Weyl fermions hosted by Dirac/Weyl cones, such as extremely high bulk carrier mobility [15], large magnetoresistance [15], and potential topological superconductivity [16], hold tremendous potential for technological applications. In 3D Dirac semimetals such as $Na_3Bi$ [1-2] and $Cd_3As_2$ [3-6], the four-fold degenerate band crossings at Dirac nodes are protected by the crystal symmetry. When the spin degeneracy is lifted by broken time-reversal or spatial inversion symmetry, a Dirac state evolves to a Weyl state where each Dirac cone splits to a pair of Weyl cones with opposite chirality[1, 3, 7-8]. The inversion symmetry broken Weyl states were demonstrated in transition metal monopnictides (Ta/Nb)(As/P) [7-14], photonic crystals [17], and $(W/Mo)Te_2$ [18-27]. The spontaneous time reversal symmetry breaking Weyl state has been reported in $YbMnBi_2$ [28] and predicted in magnetic Heusler alloys [29-33] and the magnetic members of R-Al-X (R=rare earth, X=Si, Ge) compounds [34].

In addition to the aforementioned topological semimetals with discrete Dirac/Weyl nodes, a new type of topological semimetal with Dirac bands crossing along a one-dimensional line in momentum space has also been predicted [35-43] and experimentally observed in several compounds such as $PbTaSe_2$ [44], ZrSiS [45-46], and $PtSn_4$ [47]. Among these materials, ZrSiS, which possesses layered tetragonal structure (see Fig. 1a) [43, 45, 48], shows distinct properties. Firstly, the energy bands crossing the Fermi level in ZrSiS are all Dirac bands [45], making it an ideal system to study novel physics of Dirac electron without interference of normal electrons. Secondly, this material harbors two types of unusual Dirac cones [45]: one forms a closed loop of Dirac nodes near the Fermi level in Brillouin zone and has a much wider energy range of linear dispersion (up to 2eV) compared with other known Dirac/Weyl materials. The other has an isolated Dirac node below $E_F$ near the X point in the Brillouin zone and represents the first example of a two-dimensional (2D) Dirac cone generated by a square lattice and protected by the non-symmorphic symmetry [45, 49], which does not open a gap regardless of the spin-orbit coupling strength [49]. Previous studies also showed the 2D Dirac state is hybridized with an unusual surface state [45]. These unique features make ZrSiS a particularly interesting platform for the study of novel Dirac fermion physics.

The magneto-transport and quantum oscillation measurements are important techniques for extracting the relativistic natures of Dirac fermions in topological quantum materials. High mobility and light effective mass in ZrSiS has been probed in previous Shubnikov-de Hass (SdH) oscillation studies [50-53]. However, electron transport in metals is governed by scattering mechanisms. The scattering probability varies with the number of available states that electron can be scattered into [54], therefore it oscillates in concert with the oscillations of density of state near Fermi level DOS($E_F$) and gives rise to SdH oscillations. Because SdH oscillations originate from the oscillating scattering rate, it can be complicated by the detailed scattering mechanisms, particularly in materials with lower dimensionality. For example, in some layered organic conductors, oscillations of the interlayer velocity in the SdH effect can interfere with the DOS($E_F$) oscillation, which leads the observed resistivity oscillations to substantially deviate from the prediction by the standard quantum oscillation theory, *i.e.* the Lifshitz-Kosevich (LK) theory (see Ref [55] and references therein). In contrast, the de Haas-van Alphen (dHvA) effect can always be well fitted to the LK model [55], since the oscillating magnetization directly originates from the oscillations of electrons' free energy. As a result, the dHvA effects can provide more direct information on Fermi surface (FS), particularly for those materials that are not exactly three dimensional (3D), such as ZrSiS which possesses a layered structure (Fig. 1a).

In this work, we report the first observation of dHvA quantum oscillations in ZrSiS single crystals under low magnetic fields, from which we have found important Dirac fermion properties that were not revealed in previous SdH studies [50-52]. In contrast with the SdH oscillations which attenuate quickly when the field is rotated close to the crystallographic *ab*-plane, the dHvA oscillations in ZrSiS were found to be strong under arbitrary field orientations. By taking advantage of this, we were able to investigate anisotropic behavior of Dirac cone states in ZrSiS and found that the FS enclosing the Dirac nodal line is highly anisotropic and of significant 3D character. In addition, we observed a very small Fermi pocket of strong 2D characteristic. The Dirac fermions hosted by this pocket exhibit distinct properties, including nearly zero effective mass (~$0.025m_0$) and exceptionally high quantum mobility ($10^4$ cm$^2$/Vs). These superior Dirac fermion properties, combined with a large Landé *g*-factor (~38), result in surprisingly strong Zeeman splitting of Landau levels (LLs) at low magnetic fields. These new findings pave a way for further understanding novel physics of Dirac fermions in ZrSiS.

## II. Experiments

The ZrSiS single crystals were prepared by a chemical vapor transport method. The stoichiometric mixture of Zr, Si, and S powder was sealed in a quartz tube with iodine used as a transport agent (2 mg/cm$^3$). Plate-like single crystals with metallic luster (Fig. 1a, inset) were obtained via the vapor transport growth with a temperature gradient from 950 °C to 850 °C. The composition and structure of ZrSiS single crystals were checked by using X-ray diffraction and an energy-dispersive X-ray spectrometer. The magnetization measurements were performed by a 7T SQUID magnetometer (Quantum Design). The angular-dependence of dHvA oscillations were measured using a home-made sample holder. The magnetotransport measurements, including resistivity and Hall effect measurements, were performed using a standard four- and five- probe technique in a Physics Properties Measurement System (PPMS, Quantum Design).

### III. dHvA oscillations in ZrSiS

In Fig. 1b, we present the isothermal out-of-plane ($B//c$) magnetization measured up to 7T of a ZrSiS single crystal, which exhibits strong dHvA oscillations superimposed on a paramagnetic background. The oscillations become much more visible after removing the background (Fig. 1c); they extend to a field as low as ~ 0.7 T near 2K and can sustain up to 20K. From Fig. 1b, it can be seen clearly that the oscillations consist of two components with distinct frequencies. The higher frequency component, obtained by filtering the lower frequency component, is shown in Fig. 1d. From the Fast Fourier transform (FFT) analyses of the oscillatory magnetization $\Delta M$, we have derived two oscillation frequencies which are 8.4 T ($F_\alpha$) and 240 T ($F_\beta$) respectively (Fig. 2a). We note similar low and high frequencies have also been observed in previous SdH [50-53] and thermoelectric oscillation studies [56]. As will be discussed later, both the low and high frequency oscillations originate from the Fermi surface (FS) comprised of the nodal-line bands, The extreme FS cross section areas $A_F$ estimated using the Onsager relation $F = (\Phi_0/2\pi^2)A_F$ ($\Phi_0 = h/2e$ is the magnetic flux quantum) are ~0.08 and 2.3nm$^{-2}$, respectively, for the Fermi surface pocket associated with $F_\alpha$ and $F_\beta$. Additionally, we also note that the $F_\alpha$ component exhibits a remarkable 2$^{nd}$ harmonic peak in its FFT spectrum (Fig. 2a), while the $F_\beta$ component does not. The $F_\alpha$ component also has a much greater oscillation amplitude than the $F_\beta$ component (see Fig. 1c). A large quantum oscillation amplitude and strong high harmonic components are generally expected for 2D/quasi-2D band structures [55]. Indeed, the angular dependences of $F_\alpha$ and $F_\beta$ presented below have revealed that the FS comprised of the $F_\beta$ band is of 3D nature, whereas the FS comprised of the $F_\alpha$ band is of remarkable quasi-2D character though overall it is 3D.

For the $F_\alpha$ component, the dHvA oscillations exhibit a surprisingly strong Zeeman effect at low fields, which manifests itself as broadening and splitting in the oscillation peaks, as shown in the upper inset to Fig. 1c where the split peaks are indicated by arrows. With increasing temperature from 1.8K to 12K, the split peaks gradually merge into a single one, consistent with the general expectation for the Zeeman effect, *i.e.* the thermal broadening of LLs would smear out the Zeeman splitting. The strong Zeeman splitting of oscillation peaks is shown more clearly in the susceptibility d*M*/d*B* oscillations, as seen in the lower inset of Fig. 1c where the high frequency oscillation component ($F_\beta$) has been filtered out for clarity. The threshold field for discernible peak splitting is as low as 1.7 T, which, to the best of our knowledge, is the smallest among all known topological semimetals (*e.g.*, $B > 17T$ for $Cd_3As_2$ [57-59] and TaP [60]). Such surprisingly strong Zeeman splitting reflects the superior topological Dirac fermion properties, including nearly zero effective mass and exceptionally high quantum mobility. The value of the Landé *g*-factor can be evaluated from the peak splitting. In the case of the Zeeman energy being smaller than the half of LL spacing [54], *g* is estimated to be ~38 using $F \cdot (\frac{1}{B^+} - \frac{1}{B^-}) = \frac{1}{2} g \frac{m^*}{m_0}$ [54], where $\frac{1}{B^\pm}$ represents the inverse fields of the split peaks, $\frac{m^*}{m_0}$ is the ratio of effective mass and free electron mass (~0.025, see below), and *F* is the quantum oscillation frequency. Alternatively, from the spin-zero method which is frequently used in quantum oscillation studies of organic conductors [55] and High-$T_c$ cuprates [61-63], a similar g-factor (g=37.4) can also be obtained from the angular dependence of the dHvA oscillation amplitude (see Supplementary Material for the discussions of spin-zero method).

The nature of Dirac electrons participating in quantum oscillations can be revealed from further quantitative analyses of dHvA oscillations. If higher harmonic frequency is not significant, the oscillatory magnetization can be described using the 3D Lifshitz-Kosevich (LK) formula [54, 64] which takes Berry phase into account for a Dirac system [65]:

$$\Delta M \propto -B^{1/2} R_T R_D R_S \sin[2\pi(\frac{F}{B}-\gamma-\delta)] \qquad (1)$$

where $R_T = \alpha T\mu/[B\sinh(\alpha T\mu/B)]$, $R_D = \exp(-\alpha T_D\mu/B)$, $R_S = \cos(\pi g\mu/2)$, and $\mu=m^*/m_0$. $T_D$ is Dingle temperature, and $\alpha = (2\pi^2 k_B m_0)/(\hbar e)$. Both the thermal and Dingle damping factors, *i.e.* $R_T$ and $R_D$, are due to LL broadening, caused by the finite temperature effect on the Fermi-Dirac distribution and the electron scattering respectively. The spin reduction factor $R_S$ due to Zeeman splitting, however, is field-independent and cannot describe any peak splitting in Eq. 1. To reproduce the spin-splitting of oscillation peaks, a more complicated LK formula with higher harmonic terms being included is necessary [54]. The oscillation of $\Delta M$ is described by the sine term with the phase factor $-\gamma-\delta$, in which $\gamma = \frac{1}{2}-\frac{\phi_B}{2\pi}$ and $\phi_B$ is Berry phase. The phase shift $\delta$, which is determined by the dimensionality of FS, is 0 and ±1/8, respectively for 2D and 3D cases. For the 3D case, the sign of $\delta$ depends on whether the probed extreme cross-section area of the FS is maximal (-) or minimal (+) [64]. We will show below that our observed dHvA oscillations can be well fitted to the above LK formula.

The fit of the temperature dependence of the FFT amplitude to the thermal damping factor $R_T$ (Fig. 2b) yields an effective cyclotron mass of $m^*_\alpha = 0.025 m_0$ for the low-frequency component $F_\alpha$ and $m^*_\beta = 0.052\ m_0$ for the high-frequency component $F_\beta$, much lower than those obtained from the SdH oscillations (0.1~0.27 $m_0$) [50-52]. Such a discrepancy is likely due to the fact that the LK formula cannot always precisely describe the SdH oscillations in low layered materials [55] as indicated above. Furthermore, for the $F_\alpha$-band, the previous analyses of SdH oscillations did not take the Zeeman splitting effect into account [50-52], as will be discussed later. The inverse magnetic field $1/B$ used in the $R_T$ fit is the average inverse field used for FFT analysis. The nearly zero effective mass for the $F_\alpha$-band derived from our dHvA analyses is comparable with that of the gapless Dirac semimetal $Cd_3As_2$ [15, 57-59, 66]. By fitting the field dependence of the oscillation amplitude normalized by $R_T$ to $R_D$ (Fig. 2c), the Dingle temperature $T_D$ was determined to be 8.8 K for the $F_\alpha$ band at 1.8K (note that due to the strong Zeeman splitting of oscillation peaks, the oscillation amplitude used in fitting is taken from the oscillation minima). The quantum relaxation time $\tau_q\ [= \hbar/(2\pi k_B T_D)]$ corresponding to $T_D = 8.8$K is $1.4 \times 10^{-13}$ s, from which the quantum mobility $\mu_q\ [= e\tau/m^*_\alpha]$ is estimated to be 10000 $cm^2$/Vs. Increasing temperature leads to an enhanced scattering rate, which gives rise to increased $T_D$ (Fig. 2c) and suppressed $\mu_q$; *e.g.* $\mu_q$ estimated from the Dingle plots (Fig. 2c) for 5 K and 10 K drops to 5600 and 3255 $cm^2$/Vs respectively. For the $F_\beta$ oscillation component without showing Zeeman splitting effect, the oscillation pattern can be fitted directly using the LK formula (Eq. 1) with the fixed parameters of an effective mass (0.052$m_0$) and frequency (240T). As shown in Fig. 1d, the LK formula reproduces the oscillations at $T$=2K very well, yielding $T_D$ ~6 K and a mobility of 6868 cm$^2$/Vs. The extracted quantum mobilities for both types of Dirac states are much higher than those obtained from the SdH oscillations ($\mu_q$=1~6×10$^3$ cm$^2$/Vs) [50, 52], but are comparable to those of Dirac semimetals Cd$_3$As$_2$ [57] and Na$_3$Bi [67]. The underestimated $\mu_q$ from the SdH analyses may be caused by the Zeeman effect not being taken into consideration and the possible breakdown of the LK formula for the SdH oscillations in layered materials as indicated above.

In addition to light effective mass and high mobility, the Berry phase of π accumulated along cyclotron orbits is another key feature of Dirac fermions. The Berry phase is manifested in the phase shift in quantum oscillations, which can be determined either directly from the fit to the LK formula (Eq. 1) or the LL index fan diagram. From the LK-fit of oscillations with $F_\beta$ = 240T (Fig. 1c), a phase factor of $-\gamma-\delta$=1.21 is obtained, from which the Berry phase $\phi_B$ is determined to be -0.83π ($\delta$ = -1/8) or -0.33π ($\delta$ = 1/8), as shown in Table 1, which is clearly non-trivial. This is consistent with the previous SdH [50-51, 53] and thermoelectric [56] quantum oscillation studies from which non-trivial Berry phase for the high frequency band has also been obtained by fitting the oscillation patterns or by using the LL fan diagram (*i.e.* the LL index $n$ as a function of the inverse of magnetic field 1/$B_n$). As for the oscillations with $F_\alpha$ = 8.4T, a direct fit to the above LK formula is impossible due to strong Zeeman splitting, but the Berry phase can be evaluated using the LL fan diagram. According to customary practice, the integer LL indices $n$ should be assigned when the Fermi level lies between two adjacent LLs [68-69], where the density of state near the Fermi level DOS($E_F$) reaches a minimum. Given that the oscillatory magnetic susceptibility is proportional to the oscillatory DOS($E_F$) (*i.e.* d(Δ$M$)/d$B$ ∝ Δ(DOS($E_F$)) and that the minima of Δ$M$ and d(Δ$M$)/d$B$ are shifted by π/2, the minima of Δ$M$ should be assigned to $n$-1/4. The established LL fan diagram based on this definition is shown in Fig. 2d. The extrapolation of the linear fit in the fan diagram yields an intercept $n_0 \approx 0.63$, which appears to correspond to a Berry phase of $\phi_B=2\pi(0.63+\delta)$ for the $F_\alpha$-band. However, from a careful inspection of the LL index plot in Fig. 2d, one can see a slight deviation from linear-dependence at high field (*i.e.* low 1/$B$), which can be attributed to the strong Zeeman effect as previously predicted [70]. The large Zeeman energy arising from the large $g$-factor (~38) causes the FS to be strongly field-dependent. Particularly, for a small FS pocket like the $\alpha$-FS in ZrSiS, a Lifshitz transition associated with the suppression of the spin-down FS can be expected at a moderate field. Therefore, a more reliable Berry phase might be obtained from the low field oscillation data. Indeed, we found $n_0$ decreases when the field range used for the fit shrinks toward the low field end (inset to Fig. 2d); $n_0$ drops to 0.34 when we used only the oscillation data below 1T for the linear fit, as shown by the red fit line in Fig. 2d. These results suggest that the actual Berry phase accumulated in the cyclotron orbits should be close to $2\pi(0.34+\delta)$ for the $F_\alpha$ band (*i.e.* 0.43$\pi$ for $\delta = $ -1/8 or 0.93$\pi$ for $\delta = $ 1/8, as listed table 1). Of course, the fit made below 1T covers limited data points, which may give rise to larger uncertainty for the fitted Berry phase. The nontrivial Berry phase for the low frequency (8.4T) band obtained from our dHvA oscillation studies is in contrast with that obtained from the thermoelectric oscillations, in which the intercept of the LL index plot is nearly zero [56]. Such inconsistency may be caused by the fact that the integer LL index was assigned when DOS($E_F$) reaches a minimum in our work but maximum (*i.e.*, maximum thermopower) in Ref. [56].

In order to further examine the dimensionality of Dirac cones in ZrSiS, we measured the angular dependence of dHvA oscillations and observed strong oscillations for any field orientations. This contrasts sharply with the SdH oscillations, which became hardly observable when the magnetic field was rotated close to the *ab*-plane [51-52]. Such inconsistency between SdH and dHvA effects, which is often observed in materials with low dimensionality [55], is possibly caused by the anisotropic scattering rate, which is expected for ZrSiS due to its layered structure. Given that the SdH effect is dependent on specific scattering mechanism as noted above, the higher scattering rate along the interlayer direction is likely responsible for the suppression of SdH oscillations for *B//ab*. Strong dHvA oscillations for any field orientation provide us with an opportunity to investigate the anisotropic Dirac fermion properties of ZrSiS. In Fig, 3a and 3b, we show dHvA oscillations superimposed on a diamagnetic background and their oscillatory components, respectively, for *B//ab*. The diamagnetic background seen here as well as the paramagnetic background seen in the out-of-plane magnetization measurements (Fig. 1b) should be attributed to the different sample holders used for these two sets of measurements. Compared with the dHvA oscillations of *B//c* which exhibit only two frequencies ($F_\alpha$ = 8.4 T and $F_\beta$ = 240T), the dHvA oscillations of *B//ab* consist of more frequencies. Figs. 3e and 3f display separated low and high frequency components respectively. Both components display clear beat patterns, indicating that both contain multiple frequencies. Indeed, FFT analyses (Fig. 3c) reveal two small frequencies ($F'_{\alpha 1}$ =17.6T and $F'_{\alpha 2}$ =24.5T) and three large frequencies ($F_{\eta 1}$=168T, $F_{\eta 2}$=171T, and $F_{\eta 3}$=181T). Similar multi-frequency oscillations for *B//ab* have also been observed in the isostructural compound ZrSiSe [71]. Both ZrSiS and ZrSiSe, as well as recently-reported Dirac material ZrGeM [72], ZrSnTe [73], and HfSiS [74-76], belong to a larger family of materials *WHM* with the PbFCl-type structure (*W*=Zr/Hf/La, *H*=Si/Ge/Sn/Sb, *M*=O/S/Se/Te) [43].

Similar to the in-plane cyclotron motions for *B//c*, electrons participating in the interlayer cyclotron motions for *B//ab* are also Dirac electrons, featuring light effective mass, high quantum mobility, and a non-trivial Berry phase. Compared with the case of *B//c*, electron cyclotron masses for *B//ab* (Fig. 3d) are slightly larger and in the range of 0.027-0.068$m_0$ for all probed oscillation frequencies (Table 1), whereas the quantum mobility is reduced significantly. Due to the presence of beat patterns in both low and high frequency oscillation components (Figs. 3e and 3f), quantum mobility cannot be obtained through the conventional Dingle plot, but can be through the direct fit of the oscillation pattern to the LK formula (Eq.1). The fits were performed using the multiband LK formula for which the multiple-frequency oscillations are treated as linear superposition of several single-frequency oscillations. Such an approach has been shown to be effective for analyzing the SdH oscillations of multiband Weyl semimetal TaP [60] and the dHvA oscillations of the isostructural ZrSiSe/ZrSiTe [71]. Here we fit the low- and high-frequency oscillation components separately to reduce the number of fitting parameters. As seen in Figs. 3e and 3f, with the effective mass and frequency as the known parameters, the multiband LK model reproduces the oscillation patterns very well. From the obtained Dingle temperatures of 4-16K for various frequencies (see Table 1), we have derived quantum mobilities ranging from ~2000 to ~7000 cm$^2$/Vs, which is considerably lower than that of the in-plane electron cyclotron motions for *B//c* (Table 1). Such anisotropic quantum mobility is consistent with the layered structure of ZrSiS (Fig. 1a). From the LK fits, we also obtained a non-trivial Berry phase for each band as listed in Table 1. We cannot use the widely adopted LL fan diagram method in this case, because the precise determination of the LL index field for each frequency is difficult for such multi-frequency oscillations [60].

The anisotropic characteristics of the electronic structure of ZrSiS are further clarified by systematic magnetization measurements with the magnetic field being rotated from the out-of-plane (*B//c*) to the in-plane (*B//ab*) direction (Fig. 4b). As shown in Fig. 4a, after the background subtraction, the oscillation pattern of *ΔM* displays a clear evolution with the rotation of the magnetic field. From the FFT analyses (see Supplemental Material), we have determined the angular dependence of the oscillation frequencies, as shown in Fig. 4c. The lower frequency component (shown by the black data points in Fig. 4c) exhibits a smooth evolution from *B//c* to *B//ab* though it bifurcates to two slightly different frequencies for $47° ≤ θ ≤ 90°$. The higher frequency component (shown by the blue data points in Fig. 4c) also exhibits a systematic angular evolution and splitting for $47° < θ ≤ 90°$, but its magnitude of splitting is much more significant than the lower frequency bifurcation. The $F_η$ branch, which bifurcates from the $F_β$ branch near $θ = 47°$, shows very weak angular dependence and further small splitting, while $F_β$ continues to increase until it disappears for $θ > 75°$. These angular dependences of dHvA oscillation frequencies clearly indicate that the overall FS morphology of ZrSiS has complex 3D character despite its layered crystal structure. Such observations are consistent with previous theoretical studies, which found that Zr atoms in a S-Zr-Si-Zr-S slab (see Fig. 1a) are not only bonded to neighboring S and Si layers within the slab, but also bonded to S atoms in the adjacent slabs (see Fig. 1a) [43, 48]. The 3D FS is also likely the cause of the "butterfly-shaped" anisotropic angular-dependence of magnetoresistance [51-52, 71] seen in ZrSiS. By substituting Te for S, thereby reducing the dimensionality [43], such "butterfly-shaped" anisotropy evolves to a simple two-fold anisotropy expected for a 2D system [71].

To gain more quantitative information on the Fermi surface pocket associated with the $F_β$ band, we have fitted the angular-dependence data $F_β(θ)$ to $F = F_{3D}+F_{2D}/\cos θ$ where both the 2D and isotropic 3D components are considered. We find that $F_β(θ)$ at lower angles (< 62˚) can be well fitted (Fig. 4c, red line), with the relative weight $F_{2D}/F_{3D}$ being ~1.3, suggesting dimensionality between 2D and 3D. Taking the whole angular range (0° < $\theta$ < 90°) into consideration, the continuous evolution of $F_\beta(\theta)$ and remarkable bifurcation implies a complicated 3D FS morphology. Such 3D signature is in agreement with the calculated 3D Fermi surface enclosing the Dirac nodal-line [53, 77]. Indeed, the extreme cross-section area of the FS corresponding to $F_\beta$ for $B//c$ (2.3nm$^{-2}$) matches well with the size of the Fermi pocket enclosing Dirac nodal line near X point seen in ARPES measurements [45]. A similar result was also obtained in recent high field SdH oscillation studies and first principle calculations [53], in which the 240T frequency ($F_\beta$) band is attributed to the petal-like Fermi surface pocket near R point (overlapped with the X point when projected to the $k_x$-$k_y$ plane in ARPES observations) of the Brillouin zone.

For the $F_\alpha$ band, although SdH studies have shown a field orientation-independent oscillation frequency [51], we found that $F_\alpha(\theta)$ at lower angles (< 62°) can be fitted to $F_{2D}/\cos\theta$ without considering a 3D term (Fig. 4c). At high incline angles, $F_\alpha(\theta)$ continues to evolve and persist for $B//ab$. These results imply that the $\alpha$-FS is 3D in nature, but very anisotropic, exhibiting significant 2D character. The 2D feature of the $F_\alpha$ band is also manifested in the presence of a strong harmonic peak in the FFT spectrum [55] as indicated above (Fig. 2a). Such 2D characteristics are reminiscent of the surface state observed in the ARPES experiments, which is hybridized with the non-symmorphic symmetry-protected Dirac state [45]. However, given its large oscillation amplitude (up to 7 emu/mol near $B$ = 4T, see Fig. 1c) and overall 3D nature, the $F_\alpha$-band probed in our dHvA experiments should be from a bulk state, rather than surface states. Given the nearly zero effective mass and ultra-high quantum mobility (Table 1), one might ascribe the $F_\alpha$ band to the non-symmorphic 2D Dirac band in ZrSiS, which are expected to display superior Dirac fermion properties due to gapless Dirac crossings protected by the non-symmorphic symmetry [45, 49]. However, though ZrSiS has been known as be the first material hosting such a non-symmorphic Dirac state [45], the non-symmorphic Dirac node is ~ 0.5 eV below the Fermi level, and thus is not expected to generate observable effects in quantum oscillations. Furthermore, the first principle calculations show that the entire Fermi surface of ZrSiS is comprised of only the nodal-line Dirac bands [53, 77]. As a result, the $F_\alpha$ band should also be attributed to the nodal-line Dirac band. The calculated FS for ZrSiS seems highly anisotropic and the neck of the vertical leg of the FS enclosing the nodal loop is comprised of 2D-like bands [53, 77]. Given that the cross-section areas of those necks are very small, our observed low frequency quantum oscillation component ($F_\alpha$) most likely originates from this section of Fermi surface.

## IV. Magnetotransport in ZrSiS

We have also conducted magnetotransport measurements, which provide useful information on the electronic state of ZrSiS. Our results revealed an important, previously unnoticed characteristic associated with nodal-line Dirac states – high Dirac fermion density, which holds potential for technological applications. In general, the Dirac fermion density is expected to be small when the Dirac node is close to the Fermi level. However, compared with other Dirac semimetals with discrete Dirac nodes, nodal-line Dirac semimetals are expected to have much higher Dirac fermion density due to Dirac crossings along a line/loop [71]. In order to evaluate the Dirac fermion density of ZrSiS, we measured its longitudinal resistivity $\rho_{xx}(B)$ and Hall resistivity $\rho_{xy}(B)$. As shown in Fig. 5a, $\rho_{xy}(B)$ exhibits significant non-linearity at low temperatures, implying a multiband nature, so the carrier density of ZrSiS has to be estimated from the simultaneous fits of $\rho_{xx}(B)$ and $\rho_{xy}(B)$ to a two-band model [78]:

$$\rho_{xx} = \frac{1}{e} \frac{(n_h \mu_h + n_e \mu_e) + \mu_h \mu_e (n_h \mu_e + n_e \mu_h) B^2}{(n_h \mu_h + n_e \mu_e)^2 + \mu_h^2 \mu_e^2 (n_h - n_e)^2 B^2} \qquad (2)$$

$$\rho_{xy} = \frac{1}{e} \frac{(n_h \mu_h^2 - n_e \mu_e^2) + \mu_h^2 \mu_e^2 (n_h - n_e) B^2}{(n_h \mu_h + n_e \mu_e)^2 + \mu_h^2 \mu_e^2 (n_h - n_e)^2 B^2} B \qquad (3)$$

where $n_{e(h)}$ and $\mu_{e(h)}$ are carrier density and mobility for electrons (holes), respectively. In such a simplified model, only two bands of independent carriers with characteristic density and mobility were considered [78]. The contribution of each band to the conduction process was assumed to be additive.

As shown in Fig. 5b, the fit of $\rho_{xx}(B)$ shows a small deviation for the data taken at 2K, but looks excellent for high temperatures (*e.g.* $T$=50K). This can probably be attributed to the fact that the low temperature quantum effect is not included in the classical two-band model [78]. From these fits we have extracted high carrier density, $3.64 \times 10^{20}$ and $3.59 \times 10^{20}$ cm$^{-3}$, for electron and hole bands respectively, which are significantly higher than those of other Dirac systems such as Cd$_3$As$_2$ (~$10^{18}$ cm$^{-3}$ [15, 57-59, 66]), Na$_3$Bi (~$10^{17}$ cm$^{-3}$ [79]), topological insulators (~$10^{10-12}$ cm$^{-3}$ [69]), as well as graphene ($10^{10-12}$ cm$^{-3}$ [80-81]). Furthermore, from the two-band model fitting we also estimated a transport mobility of ~13,700 cm$^2$/Vs for holes and 12,300 cm$^2$/Vs for electrons. These values are higher than quantum mobility, since the transport relaxation time is not significantly affected by small angular scattering, while the quantum relaxation is [54].

Given the relatively small quantum oscillation frequencies in ZrSiS (*e.g.*, 8.4T and 240T for *B*//*c*), the obtained high carrier density appears to suggest that the entire Fermi surface is not probed. According to the first principle calculations [53, 77], both the electron and hole pockets of ZrSiS exhibit multiple extreme cross-sections, which is in contrast with the only two frequencies observed in dHvA oscillations for *B*//*c*. This discrepancy might be due to the limited magnetic field range ($B \lesssim 7$T) in our experiments. Indeed, though only two oscillation frequencies were probed in our dHvA experiments and several other previous low field SdH studies ($B \lesssim 9$T) [50-51], the recent thermoelectric [56] and high field ($B \lesssim 33$T) SdH [53] experiments have revealed additional frequency around 600T, which is attributed to the electron pocket [53].

## V. Discussions

Compared with other topological semimetals, ZrSiS exhibits the coexistence of two types of Dirac states, *i.e.* the 3D nodal-line Dirac state and the 2D Dirac state protected by non-symmorphic symmetry. As stated above, the non-symmorphic symmetry-protected Dirac crossing is located well below the Fermi level, thus hardly contributing to quantum oscillations and electron transport. As has been revealed by Andreas *et al* [82], the position of the non-symmorphic Dirac node in *WHM* (*W*=Zr/Hf, *H*=Si/Ge, *M*=S/Se/Te) materials is determined by the crystallographic *c*/*a* ratio. It is located well below the Fermi level in ZrSiS because of the smaller *c*/*a* ratio (~ 2.27), but right at the Fermi level in ZrSiTe due to the "right" *c*/*a* ratio (~2.57), which allows for investigating the transport properties of the Dirac fermions protected by the non-symmorphic symmetry in ZrSiTe [82]. Nevertheless, under such a circumstance, the zeroth Landau level of the relativistic fermions is pinned at the Dirac node, and other Landau levels would not pass through the Fermi level upon increasing magnetic field [83], so the quantum oscillations due to the non-symmorphic Dirac fermions were not detected in our dHvA experiments on ZrSiTe [71].

Although the non-symmorphic Dirac fermions in ZrSiS cannot be probed in low energy measurements such as quantum oscillations, the nodal-line fermions in ZrSiS also exhibit distinct properties as mentioned above, such as high Dirac fermion density, nearly zero effective mass and ultrahigh mobility. These properties may lead to enhanced electrical and thermal conductivity and other possible exotic phenomena. In addition, the surprisingly strong Zeeman splitting at low magnetic fields also distinguishes ZrSiS from other Dirac materials.

Quantum oscillation peak splitting due to the Zeeman effect is widely seen in topological semimetals such as $Cd_3As_2$ [57-59] and TaP [60], which is caused by the superposition of oscillations from the split spin-up and spin-down sub LLs. Generally, it occurs at high magnetic fields (*e.g.* above 17T for $Cd_3As_2$ [57-59] and TaP [60]) when the separation of the split energy levels ($= g\mu_B B$) exceeds the breadth of LLs ($\propto 1/\tau_q$). In ZrSiS, Zeeman peak splitting can be observed in the low- frequency oscillation component ($F_\alpha$ band) at fields as low as ~1.7T (see the bottom inset in Fig. 1c), reflecting unusual properties of the nodal-line fermions. To understand the origin of the low field Zeeman effect, we must first consider the quantum relaxation time $\tau_q$ which affects the LL breadth. As listed in Table 1, $\tau_q$ for the $F_\alpha$ band is ~0.14 ps, comparable with $\tau_q$ ~0.2 ps for the $F_\beta$ band and that of other topological semimetals [57-59]. Thus the presence of low-field Zeeman splitting only in the $F_\alpha$ band cannot be understood only in light of long $\tau_q$. On the other hand, the simple superposition of two sine wave oscillations cannot produce any peak splitting, but a modulated oscillation pattern. A necessary requirement for peak splitting is the superposition of the non-sinuous oscillations with sharp peaks.

Mathematically, such "peaky" oscillations represent harmonic frequency components. Therefore, the dHvA/SdH oscillations of spin-up and down electrons must contain harmonic components to cause Zeeman peak splitting [54]. Since the $r^{th}$ harmonic term attenuates with $1/B$ in the form of $\exp[-r2\pi^2 k_B m^*(T+T_D)/e\hbar B]$, when the exponent $r2\pi^2 k_B m^*(T+T_D)/e\hbar B$ appreciably exceeds 1, the higher harmonics are damped out and the Zeeman effect occurs only through the superposition of the simple oscillations with fundamental frequencies for spin-up and down electrons. This leads to an amplitude change as described by the spin damping factor $R_s$ in Eq. 1, rather than peak splitting [54]. In ZrSiS, for the $F_\alpha$ band, the large Landé g-factor (~38) and long $\tau_q$ (0.14ps at $T$ = 2 K) ensures well separated spin-up and down LLs. The nearly zero effective quasiparticle mass (~$0.025m_0$) causes slower damping of the higher harmonics in low fields. Moreover, the quasi-2D characteristic of the $F_\alpha$ band also makes the harmonic component stronger as indicated above. As shown in Fig. 2a, we indeed observed a strong $F_{2\alpha}$ harmonic component for the $F_\alpha$ band, which accounts for our observation of strong Zeeman splitting at low fields in the dHvA oscillations of the $F_\alpha$ band. For the $F_\beta$ band, the LLs should also be sharp given its longer $\tau_q$, but Zeeman splitting is absent below 7T for this band. This can probably be attributed to the fact that the $F_\beta$ band is of significant 3D character and the effective mass ($0.052m_0$) is twice as large as that of the $F_\alpha$ band ($0.025m_0$), which causes quicker damping of the higher harmonics.

In addition to rich harmonics, a large g-factor (~38) is also needed to observe strong Zeeman splitting in ZrSiS. The spin-orbit coupling introduces an interaction energy of $\lambda \boldsymbol{L} \cdot \boldsymbol{S}$, leading to an effective g-factor of $g_{eff} = g (1 \pm \lambda \boldsymbol{L} \cdot \boldsymbol{S}/\Delta)$ where $\Delta$ is the crystal field splitting and $\lambda$ is the spin-orbital coupling constant [84]. Spin-orbit coupling is predicted to play a more important role in Dirac materials [85-87]. When the spin-orbit coupling and non-relativistic approximation are taken into account, the cyclotron energy is found to be equal to the Zeeman splitting energy derived from the Dirac equation[85-87], from which the *g*-factor is determined to be $2m_0/m_D$ where $m_0$ and $m_D$ represent free electron mass and Dirac electron mass respectively. In ZrSiS, such scenario indeed leads to a large *g*-factor =$2m_0/m_D$ =80, about twice of the measured value. Furthermore, electronic correlations also result in enhanced *g*-factor, which has been widely seen in 2D electron gas systems owing to many-body exchange interactions [88-89]. In general, for a system showing Landau level quantization, the increased degeneracy of Landau levels leads to enhanced density of state near the Fermi level under high magnetic fields, which effectively amplifies the electron correlation effect [90-92]. This effect is expected to be particularly strong for the *α*-Dirac band of ZrSiS since the quantum limit of this band can be easily reached due to its low oscillation frequency ($F_\alpha$ ~8.4T). Although our SQUID magnetometer can only reach 7T, the second Landau level can be reached and our experimentally observed large *g*-factor in ZrSiS is in line with the anticipated enhanced electronic correlation effects, as seen in $ZrTe_5$ [91]. Indeed, the recent high field SdH studies has revealed usually mass enhancement at low temperatures, which seems consistent with the enhanced correlation in ZrSiS [56].

The low field Zeeman splitting, however, was not clearly observed in SdH oscillations [50-52], since the SdH oscillations are much weaker and only 1~2 oscillation periods can be seen for the $F_\alpha$ band up to *B*=9T (see Supplemental Material and Ref. [50-52]). For this reason, the split peaks in the SdH oscillations were attributed to individual LLs, which causes the aperiodic oscillation pattern, rather broad FFT peaks, higher FFT frequency (14-23T), as well as the overestimated effective mass ($m^*$=0.12-0.14$m_0$) and underestimated quantum mobility ($\mu_q$=5~6×$10^3$ cm$^2$/Vs) [50-52] for the $F_\alpha$ band. However, as shown below, from the comparison between SdH and dHvA oscillation patterns, the SdH oscillation "peaks" are clearly attributed to the Zeeman splitting.

It has been established the oscillation part of the DOS($E_F$) is in phase with the oscillatory susceptibility ($\Delta$DOS($E_F$) $\propto \Delta$(d$M$/d$B$)) [55]. In the framework of quantum oscillation theory, SdH oscillations are interpreted as the scattering rate oscillations $\Delta(1/\tau)$ which arise from the oscillation of DOS($E_F$), since $1/\tau \propto$ DOS($E_F$) [55, 93-94]. Considering that the longitudinal resistivity $\rho_{xx}$ (see Supplemental Material for $\rho_{xx}$) is much greater than the transverse (Hall) resistivity $\rho_{xy}$ (Fig. 4e) in ZrSiS, the matrix conversion of resistivity and conductivity [$\sigma_{xx} = \rho_{xx}/(\rho_{xx}^2+\rho_{xy}^2)$] yields $\rho_{xx} \approx 1/\sigma_{xx} \propto 1/\tau$. Therefore, $\Delta\rho_{xx} \propto \Delta(1/\tau) \propto \Delta$DOS($E_F$) $\propto \Delta$(d$M$/d$B$). i.e., $\Delta\rho_{xx}$ and $\Delta$(d$M$/d$B$) are in phase.

Indeed, as shown in Fig. 6, after filtering out the high frequency ($F_\beta$) oscillation component, the low frequency ($F_\alpha$) susceptibility oscillations $\frac{d\Delta M}{dB}$ (Fig. 6a) is *in phase* with the low frequency component of the $\rho_{xx}$ oscillations (Fig. 6c). Given such an agreement, the oscillation minima in both $\frac{d\Delta M}{dB}$ and $\rho_{xx}$, which was previously thought to originate from individual Landau levels in SdH effect studies [50-52], should correspond to the split sub energy levels due to the Zeeman effect as revealed in our dHvA effect studies.

### VI. Conclusion

In summary, we have observed very strong low field dHvA oscillations due to the Dirac nodal-line fermions in ZrSiS. The analyses of our experimental data reveal the anisotropic Dirac bands hosting nearly massless Dirac fermions in ZrSiS . Furthermore, we have observed extremely strong Zeeman splitting with a large g-factor ~ 38. These findings suggest that ZrSiS is unique topological material for seeking and understanding exotic phenomena of Dirac nodal-line fermions.

**Acknowledgement**

This work was supported by the US Department of Energy under grant DE-SC0014208. The authors thank K.W. Chen at National High Magnetic Field Lab (Tallahassee) for informative discussions.

# Figures

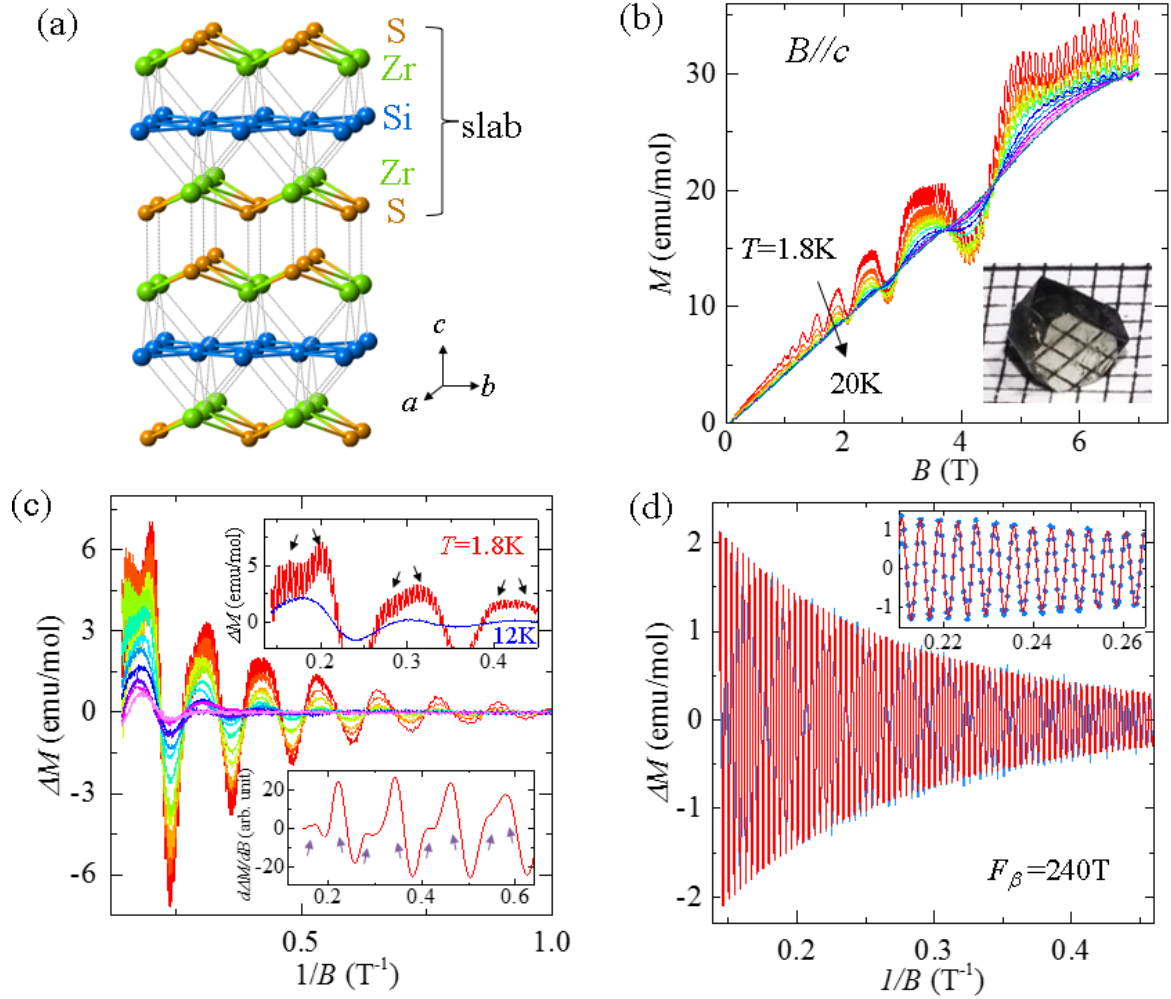

FIG. 1. (a) Crystal structure of ZrSiS. (b) Isothermal out-of-plane magnetization ($M$) for ZrSiS at various temperatures from $T$=1.8K to 20K. Inset: an image of a ZrSiS single crystal. (c) The oscillatory component of the magnetization $\Delta M$. Upper inset: Zeeman effect at $T$ = 1.8 K and 12 K. Lower inset: susceptibility oscillation d$\Delta M$/d$B$. The high frequency component is filtered out for clarity. Peak splitting is indicated by arrows and discernible above $B$=1.7 T. (d) the high frequency oscillation component (blue data points) and the fit of the oscillation pattern to the LK

formula (red solid line). To better illustrate the fitting quality, the zoomed-in data and fitting are shown in the inset.

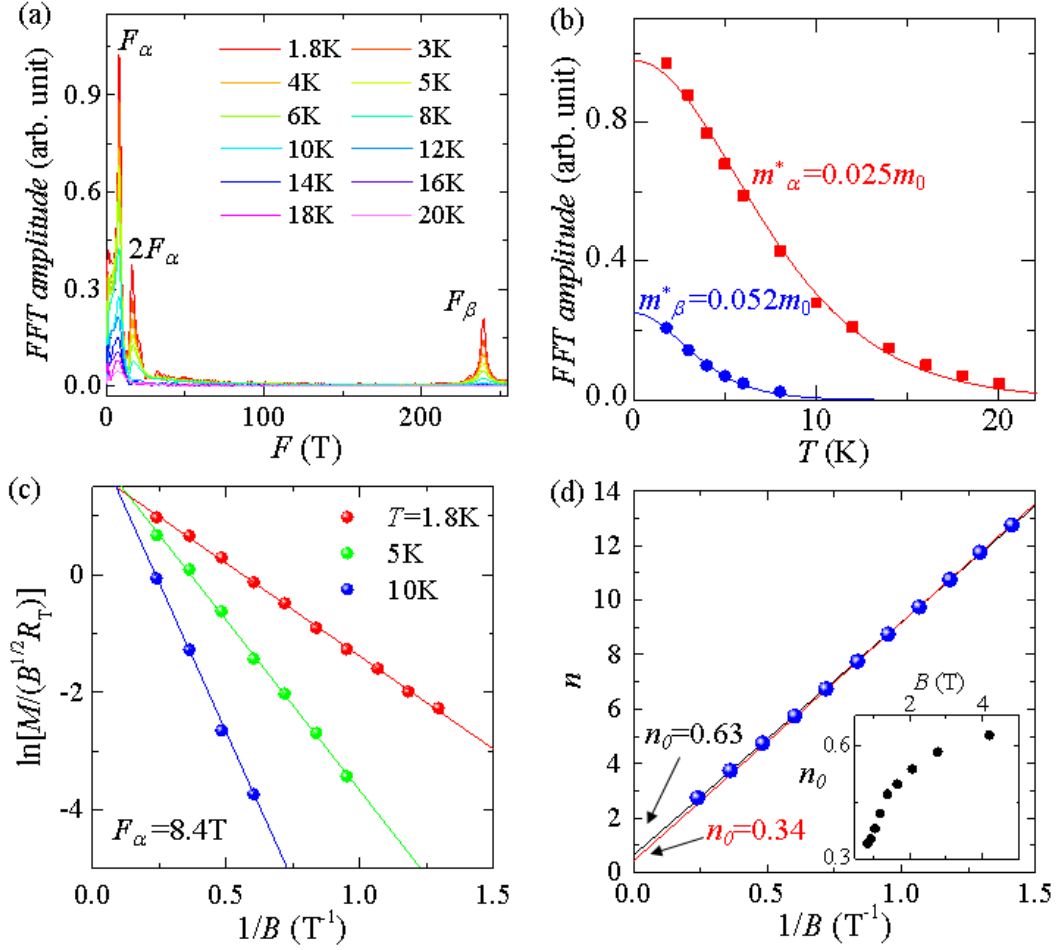

FIG.2. Analyses of the dHvA oscillations for $B//c$. (a) the FFT spectra of the oscillatory component of the dHvA oscillations for $B//c$ at various temperatures. (b) the fits of the FFT amplitudes of $F_\alpha$ and $F_\beta$ to the temperature damping factor $R_T$ of the LK formula. (c) Dingle plot for the low frequency oscillation component ($F_\alpha$ =8.4 T) at $T$=1.8K, 5K and 10K. (d) LL index fan diagram for the low frequency oscillation component. $n$-1/4 ($n$, integer LL indices) are assigned to the $\Delta M$ oscillation minima. The black line represents a linear fit to all LL indices,

which yields an intercept of $n_0 = 0.63$. The red line represents the linear fit to the LL indices obtained below 1T to minimize the influence of the Zeeman effect, which yields an intercept of 0.34. Inset: the variation of the intercept $n_0$ with the magnetic field range used for the linear fit of the LL fan diagram.

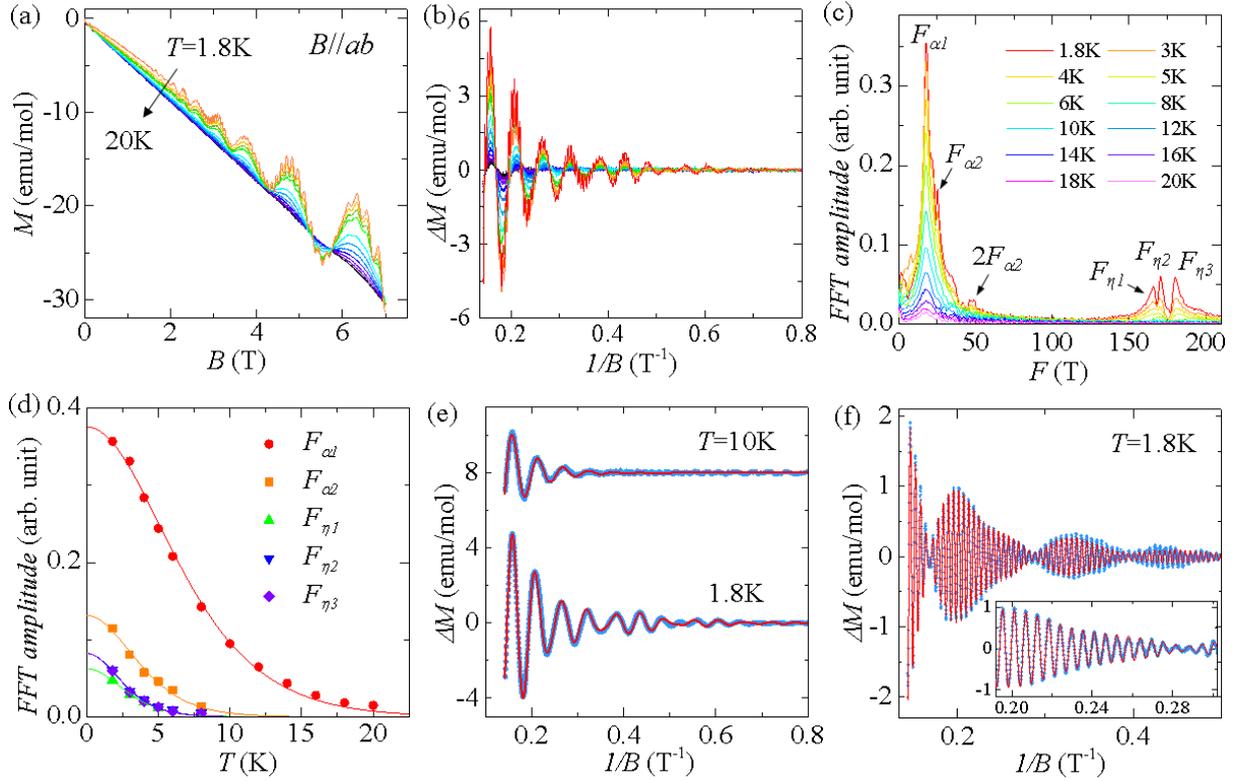

FIG. 3. Analyses of the dHvA oscillations for $B//ab$. (a) Isothermal in-plane magnetization measurements for ZrSiS at various temperatures from $T$=1.8K to 20K. (b) the oscillatory component of the magnetization $\Delta M$. (c) the FFT spectra of the oscillatory component of the in-plane magnetization at various temperatures. (c) the fits of the FFT amplitudes to the temperature damping factor $R_T$ of the LK formula. (e) and (f): the low (e) and high (f) frequency oscillation components of the dHvA oscillations. The red lines show the fits of the oscillation pattern to the generalized two-band and three-band LK formula for the low frequency (e) and high frequency

(f) oscillations. Inset in (f): the enlarged data and fit within a small field range, which clearly shows the data can be best fitted by the three-band LK formula.

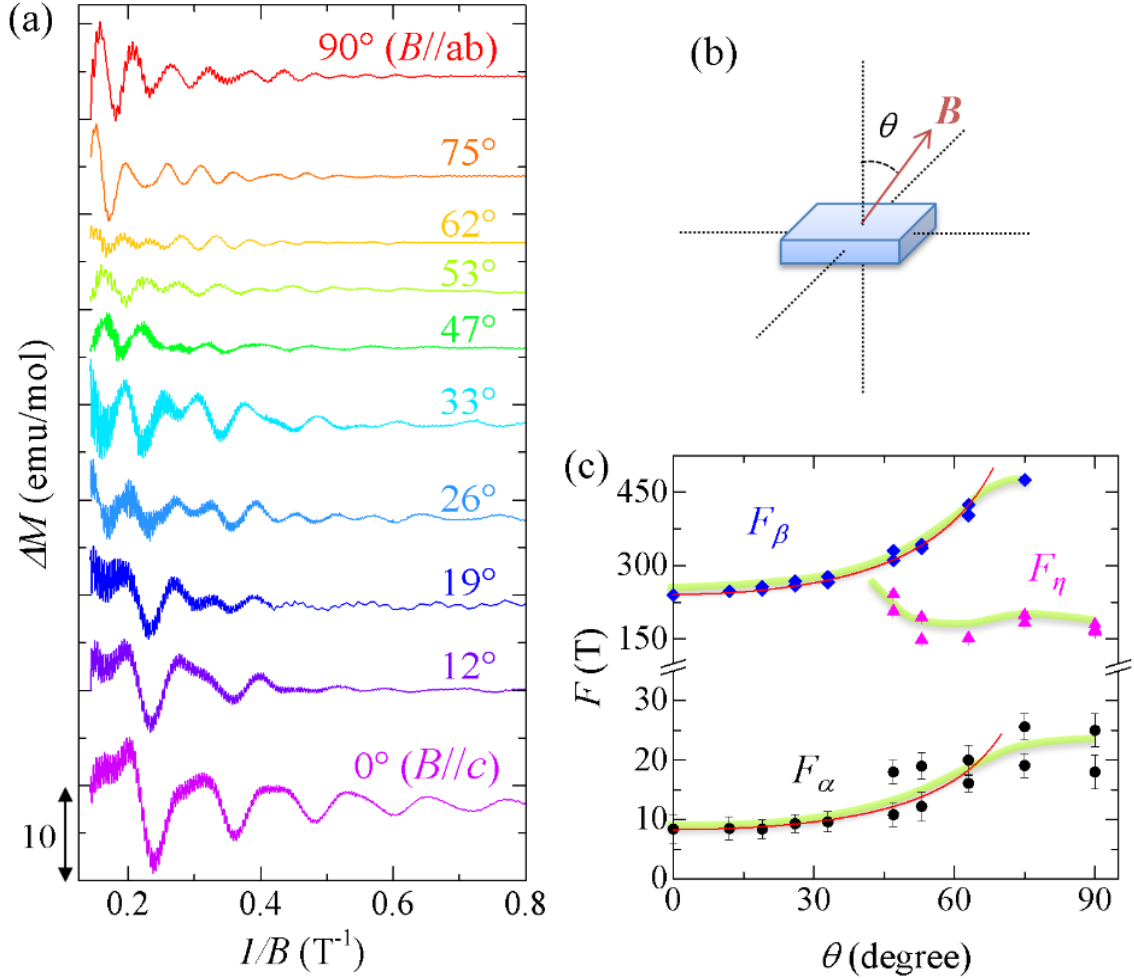

FIG. 4. Fermi surface morphology of ZrSiS. (a) dHvA oscillations at $T$=1.8K for different magnetic field orientations. Data at different field orientations have been shifted for clarity. (b) schematic of the measurement set up. The field is rotated from an out-of-plane direction ($B//c$, defined as $\theta$=0°) to an in-plane direction ($B//ab$, defined as $\theta$ =90°). (c) the angular dependence

of the oscillation frequencies obtained from the FFT spectra. Error bars are defined as the half-width at the half-height of FFT peak. The green lines are for eye guides. The red lines are fits to $F = F_{3D} + F_{2D}/\cos\theta$ for $F_\alpha$ and $F_\beta$ below 60°.

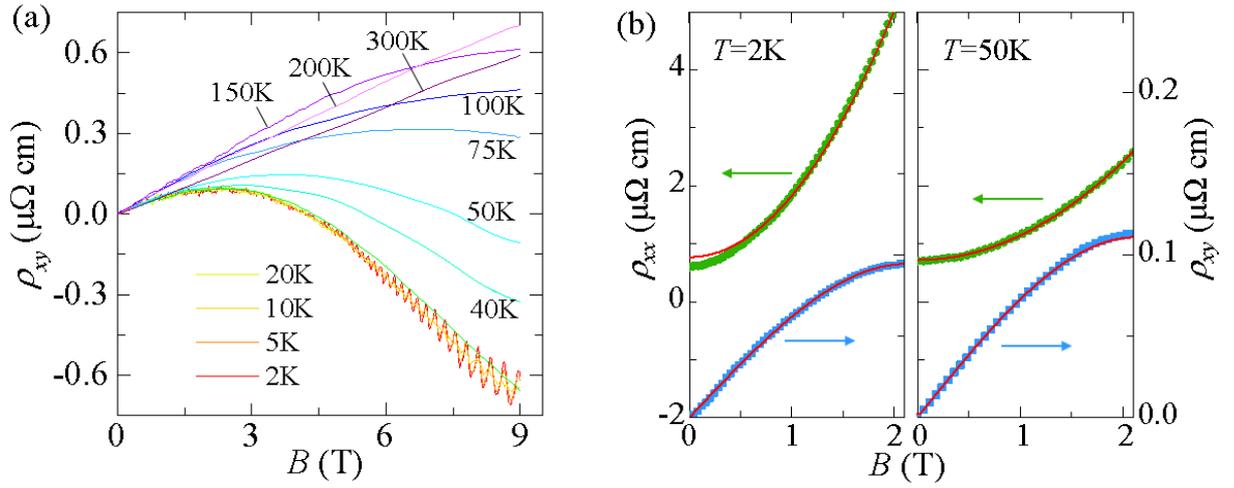

FIG. 5. Transport properties of ZrSiS. (a) Magnetic field dependence of Hall resistivity $\rho_{xy}$. SdH oscillations are seen below 20K. (b) Two-band model fitting to the transverse ($\rho_{xx}$) and Hall ($\rho_{xy}$) resistivity at $T$=2K and 50K.

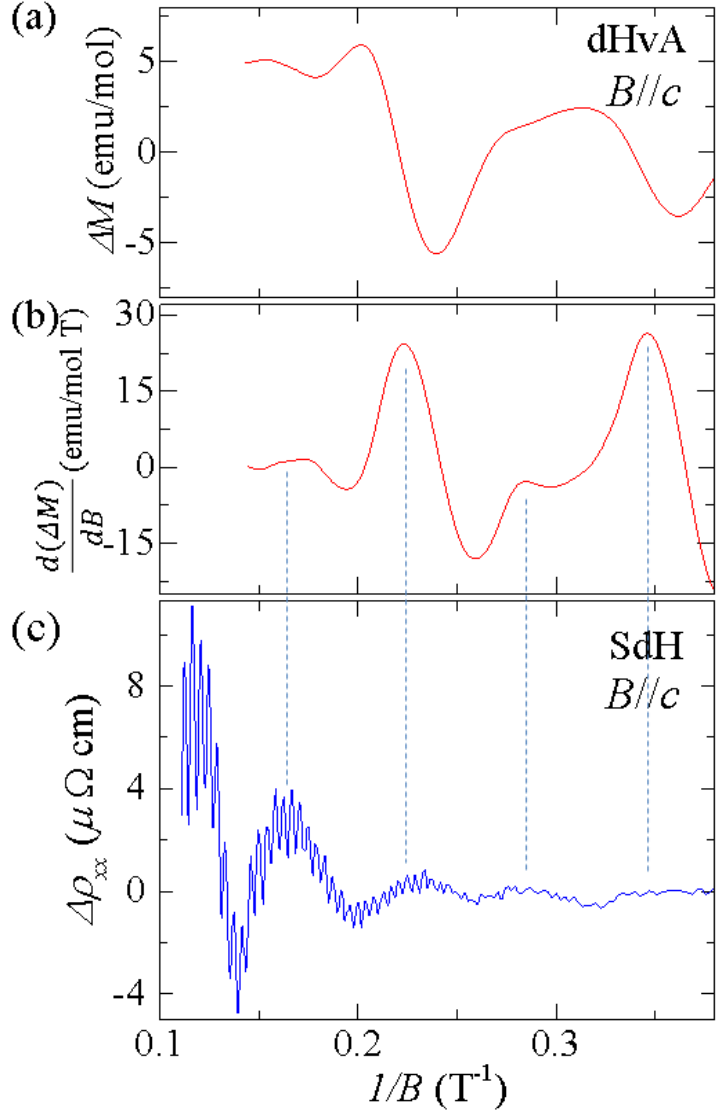

FIG. 6. Comparison of Zeeman effect in dHvA and SdH oscillations. (a) The lower frequency dHvA oscillation component of the out-of-plane ($B//c$) magnetization at $T$=2K, obtained by filtering the higher frequency component. Zeeman splitting is clearly observed. (b) Derivative of the lower frequency oscillation component, d$\Delta M$/d$B$. (c) SdH oscillations of the in-plane magnetoresistance ($\Delta \rho_{xx}$) at $T$=2K. Zeeman splitting occurs at the almost same field for both dHvA (d$M$/d$B$) and SdH ($\rho_{xx}$) oscillations, as denoted by dashed lines.

**Table 1** The oscillation frequency $F$, Dingle temperature $T_D$, effective mass $m^*/m_0$, quantum relaxation time $\tau_q$ [$= \hbar/(2\pi k_B T_D)$], quantum mobility $\mu_q$ ($= e\tau/m^*_\alpha$), and Berry phase $\phi_B$ of different Dirac bands probed by dHvA oscillations.

|  | $F$ (T) | $T_D$ (K) | $m^*/m_0$ | $\tau_q$ (ps) | $\mu_q$ (cm$^2$/Vs) | $\phi_B$ $\delta = -1/8$ | $\delta = 0$ | $\delta = 1/8$ |
|---|---|---|---|---|---|---|---|---|
| **B//c** | 8.4 | 8.8* | 0.025 | 0.14 | 10000 | 0.43π | 0.68π | 0.93π |
|  | 240 | 6** | 0.052 | 0.2 | 6868 | -0.83π | -0.58π | -0.33π |
| **B//ab** | 17.6 | 14** | 0.027 | 0.084 | 5469 | -0.14π | 0.11π | 0.36π |
|  | 24.5 | 11.3** | 0.045 | 0.108 | 4219 | -0.39π | -0.14π | 0.11π |
|  | 167.5 | 16.2** | 0.062 | 0.075 | 2127 | -0.75π | -0.5π | -0.25π |
|  | 170.6 | 4.4** | 0.066 | 0.277 | 7378 | -0.75π | -0.5π | -0.25π |
|  | 180.7 | 11.3** | 0.068 | 0.108 | 2792 | -0.75π | -0.5π | -0.25π |

(* Obtained from Dingle plot; ** Obtained from the LK fitting.)